\documentclass[12pt]{article}
\pdfoutput=1

\pdfpageattr{/Group << /S /Transparency /I true /CS /DeviceRGB>>} 
\usepackage{graphicx,epsfig}
\usepackage{amsthm}
\usepackage{amsmath}
\usepackage{amsfonts}
\usepackage{amssymb}
\usepackage{setspace}
\usepackage{color}
\usepackage{cite}
\usepackage{mcite}
\usepackage{mathrsfs}
\usepackage{enumerate}
\usepackage[utf8]{inputenc}
\newcommand{\gsim}{\lower.7ex\hbox{$\;\stackrel{\textstyle>}{\sim}\;$}}
\newcommand{\lsim}{\lower.7ex\hbox{$\;\stackrel{\textstyle<}{\sim}\;$}}

\def\mpl{M_{\rm Pl}}
\makeatletter
\newsavebox\myboxA
\newsavebox\myboxB
\newlength\mylenA

\newcommand*\xoverline[2][0.75]{%
    \sbox{\myboxA}{$\m@th#2$}%
    \setbox\myboxB\null
    \ht\myboxB=\ht\myboxA%
    \dp\myboxB=\dp\myboxA%
    \wd\myboxB=#1\wd\myboxA
    \sbox\myboxB{$\m@th\overline{\copy\myboxB}$}
    \setlength\mylenA{\the\wd\myboxA}
    \addtolength\mylenA{-\the\wd\myboxB}%
    \ifdim\wd\myboxB<\wd\myboxA%
       \rlap{\hskip 0.5\mylenA\usebox\myboxB}{\usebox\myboxA}%
    \else
        \hskip -0.5\mylenA\rlap{\usebox\myboxA}{\hskip 0.5\mylenA\usebox\myboxB}%
    \fi}
\makeatother
\def\beq{\begin{equation}}
\newcommand{\eeq}[1]{\label{#1}\end{equation}}
\def\bea{\begin{eqnarray}}
\newcommand{\eea}[1]{\label{#1}\end{eqnarray}}
\def\ba{\begin{array}}
\def\ea{\end{array}}\newcommand{\be}{\begin{equation}}
\newcommand{\ee}{\end{equation}}

\newcommand{\comment}[1]{}

\newcommand{\bsb}{\boldsymbol}

\def\ep{\epsilon}
\def\vep{\varepsilon}

\def\k{{\bsb k}}

\def\p{{\bsb p}}

\def\Q{{\mathscr Q}}

\def\S{{\mathcal{S}}}

\def\H{{\mathcal H}}

\begin{document}
\begin{titlepage}
\vspace{20pt}
\begin{center}
{\huge   Shaving off Black Hole Soft Hair}

\vspace{18pt}

{\large  Mehrdad Mirbabayi$^a$~\footnote{e-mail: mehrdadm@ias.edu}  and Massimo Porrati$^b$~\footnote{e-mail: massimo.porrati@nyu.edu}}

\vspace{12pt}

{$^a$ \em Institute for Advanced Study, \\ 1 Einstein Drive, Princeton, NJ 08540, USA}

\vspace{12pt}

{$^b$ \em Center for Cosmology and Particle Physics, \\ Department of Physics, New York University, \\4 Washington Place, New York, NY 10003, USA}

\end{center}

\vspace{12pt}

\abstract{A recent, intriguing paper by Hawking, Perry and Strominger suggests that soft photons and gravitons can be 
regarded as black hole hair {and may be relevant to the black hole information paradox.}
 In this note we make use of factorization theorems for infrared divergences of the S-matrix to argue that by appropriately 
 dressing in and out hard states, the soft-quanta-dependent part of the S-matrix becomes {essentially} trivial. The information paradox 
 can be fully formulated in terms of dressed hard states, which do not depend on soft quanta. }
\end{titlepage}
\newpage

\vspace{-1cm}
\section{Soft Hair on Black Holes}
An infinite number of asymptotic symmetries for gravity and Abelian gauge theories were uncovered in the last few year 
thanks to the work of several authors, especially A. Strominger~\cite{Strominger_graviton1,Strominger_graviton,Strominger_photon,Strominger_QED,Campiglia:2015kxa}. A recent, 
intriguing paper \cite{Hawking} by Hawking, Perry, and Strominger argues that such new symmetries can be used to  constrain the final 
states resulting from black hole evaporation \cite{Hawking74,Hawking75}, beyond the universal restrictions due to energy and charge conservation. 
{This fact is potentially relevant to the black hole information paradox~\cite{Hawking76}.} Two new 
ingredients enter in their discussion. The first one is the existence of the infinite-dimensional set of new symmetries 
mentioned above. Each symmetry generates a conserved charge. The second ingredient {involves a clever use
 of} such charges to create new black hole states out of old ones. The crucial claim of ref.~\cite{Hawking} is that these new
 states are distinguishable from the old ones. 

By itself, the existence of new conserved charges does not imply the existence of new black hole hair. In the specific case 
considered in~\cite{Hawking}, new $U(1)$ asymptotic charges are obtained by integrating a trivially conserved current, 
$J=\star d (\vep \star F)$, over an appropriate Cauchy surface. In the absence of black holes or massive charged states, 
the surface can be pushed up to future null infinity $I^+=R\times S^2$. When the scalar function $\vep$ is independent of 
the null generator of $I^+$, but has an arbitrary dependence on the angular coordinates $(z, \bar{z})$ of 
${S}^2\in I^+$, the charge is
\beq
\Q=\int _{I^+} d (\vep \star F) = \int_{I^+} \hat{d} \vep \wedge \star F +  
\int_{I^+} \vep d \star F
\eeq{m1}
The term $\Q_S\equiv \int_{I^+} \hat{d} \vep \wedge \star F$, where $\hat{d}$ is the exterior derivative on $S^2 \in I^+$, is the 
``soft charge'' of ref.~\cite{Strominger_photon}, while $\Q_H=\int_{I^+} \vep d \star F=\int_{I^+} \vep \star j$ is the hard 
charge. The last equality uses of course Maxwell's equations.

In the presence of a classical black hole, even in the simplest case that no massive charged matter exists, $I^+$ is no longer a Cauchy surface. On the other hand, a black hole hair is an object defined on $I^+$ (and not on the horizon) that can be used to reconstruct the black hole state. It is the total derivative nature of the current $J$ that makes $\Q$ a potential black hole hair. Namely, as in the case of black hole electric charge and ADM mass, $\Q$ can be written as a surface integral over the sphere at spatial infinity, or $I^+_-$,
\beq
\Q = -\lim_{u\to -\infty} \oint  \vep \star F,
\eeq{boundary}
where $u$ is the retarded time. However, for a classical stationary black hole space-time all new charges are trivial \cite{Flanagan}, as expected from black hole no-hair theorems.
\begin{figure}[h]
\begin{center}
\epsfig{file=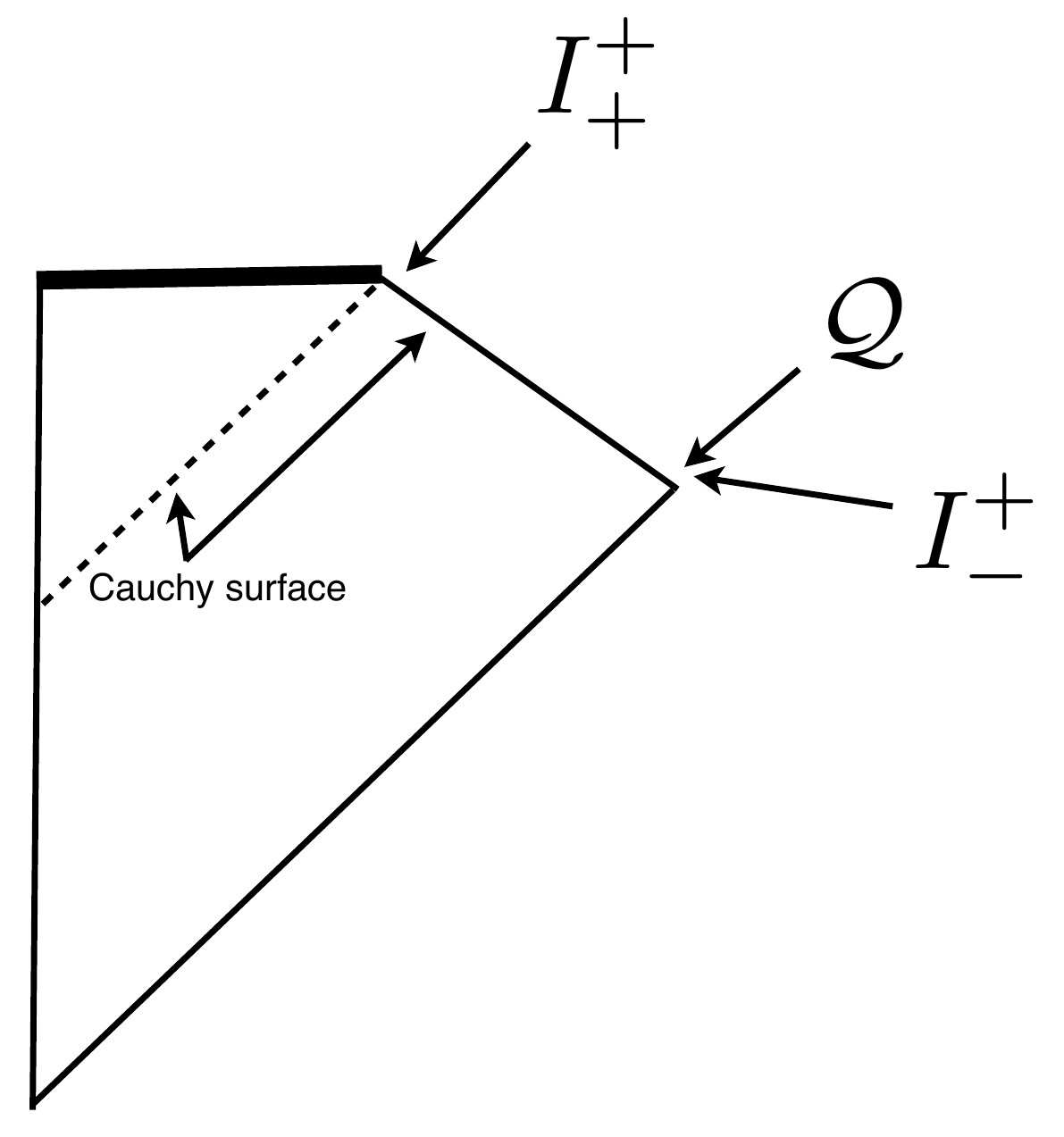, height=4in, width=3.7in}
\end{center}
\caption{In the presence of a classical black hole, the Cauchy surface is $I^+\cup H$, but the charge $\Q$ is 
an asymptotic quantity, as it can be written as a boundary integral at  $I^+_-$. However, $\Q=0$ for classical stationary black holes unless $\vep = {\rm constant}$, in which case it is a multiple of the black hole electric charge.}
\label{vaidya}
\end{figure}

Consider next a quantum black hole which evaporates. Here the second ingredient in the 
Hawking-Perry-Strominger mechanism enters and becomes essential. 
As a warm-up example to the Hawking-Perry-Strominger mechanism, let's ask a simpler question:\footnote{{We thank Dan Harlow for suggesting this analogy to us.}} is the three-momentum vector a black hole hair? Hawking, Perry, and Strominger would argue that it is \cite{Hawking}. And indeed it has an implication for Hawking evaporation. If the early Hawking quanta, from an initially stationary black hole, carry away total momentum $\bsb P$, then by momentum conservation the resulting black hole must have a nonzero momentum $-\bsb P$ and so do the late Hawking quanta. This is a source of correlation between the early and late Hawking radiation, which makes the final state less mixed than a thermal state. However, the correlation is much too small to purify the Hawking radiation.

This simple fact can be related to the existence of a symmetry operator that transforms the black hole state. Suppose 
that after the emission of early quanta the ADM mass of the black hole is $M$, and it is sufficiently large that we can talk about a metastable state $|M\rangle$ with some internal degrees of freedom, not explicitly shown in $|M\rangle$. A moving black hole state can be obtained from the stationary one, {described by} $|M\rangle$, by a boost $U(\Lambda)$, where $\Lambda(M,\bsb 0) = (\sqrt{M^2+P^2},-\bsb P)$. Lorentz symmetry implies that the S-matrix $\S$ commutes with boosts, so, if $|M\rangle$ evaporates into $\S|M\rangle\equiv |X\rangle$, then  
\be
\S U(\Lambda) |M\rangle = U(\Lambda) \S |M\rangle = U(\Lambda)|X\rangle.
\ee
{The final state} $|X\rangle$ can be expanded in terms of asymptotic states
\be
|X\rangle = \sum_{b} \S_{M\to b} |b\rangle, \qquad |b\rangle = \prod_{i=1}^m a_{p_i,\zeta_i}^\dagger |0\rangle
\ee
where $b = \{(\p_1,\zeta_1),\cdots,(\p_m,\zeta_m)\}$ runs over outgoing states and $\zeta$ characterizes their discrete quantum numbers. Applying $U(\Lambda)$, the Hawking quanta which are momentum eigenstates get boosted, while the vacuum is boost invariant
\be
U(\Lambda)|X\rangle = \sum_{b} \S_{M\to b} |\Lambda b\rangle, \qquad U(\Lambda) |0\rangle = |0\rangle.
\ee
Thus the late-time observer can distinguish $|M\rangle$ from $U(\Lambda)|M\rangle$ 
{by measuring the $a_{p_i,\zeta_i}$ quanta. Notice that these are in general ``hard," since their momenta
are generic}.

One can ask if super-translation symmetries \cite{Strominger_graviton1,Strominger_graviton} and their analog in electrodynamics \cite{Strominger_photon} (hereafter denoted as {large $U(1)$} symmetries) lead to additional hair in a similar way. Naively, given that there are infinitely many conserved charges (involving energy flux and electric charge flux in every direction) then, depending on the angular distribution of early quanta, there will {exist} very non-trivial constraints on the late quanta. This would lead to much larger correlations between late and early radiation. 

\section{Shaving off the Soft Hair}

We will now show that these conservation laws fix early (late) soft radiation in terms of early (late) hard radiation, but do not induce any cross correlation between early and late quanta. The easiest way to see this is to introduce a new basis of asymptotic states in which hard particles are dressed with soft photons and gravitons. In terms of {this new basis}, the soft part of the $\S$-matrix becomes trivial and all  conservation laws are automatically satisfied.

First, choose an IR cutoff $\lambda$, much {smaller} than the typical energy $E$ of the particles involved in the process. 
In the case of black holes, $E$ is the Hawking temperature. Write In and Out Hilbert spaces as products 
$\H^\pm = \H_h^\pm \otimes \H_s^\pm$ where $\H_s^+$ ($\H_s^-$) includes soft outgoing (incoming) photons and 
gravitons with frequency less than $\lambda$. Any In state can be written as a superposition of states of the form 
$|a\rangle |\alpha\rangle$, where $a \in \H^-_h$ labels the momenta and quantum numbers of hard In states and 
$\alpha\in \H^-_s$ labels soft incoming photons/gravitons. Every Out state is similarly written as $|b\rangle |\beta\rangle$. 

The Weinberg soft theorems \cite{Weinberg,Weinberg-I} imply that, for fixed initial ($|a\rangle$) and final ($|b\rangle$) hard
 states, the S-matrix matrix factorizes into the product of: 
\begin{enumerate}
\item A  ``hard'' unitary matrix, $\hat{\S}$, 
which does not depend on soft degrees of freedom. This means that $\hat{\S}$ 
acts as the unit matrix on the space of soft photons/gravitons. 
\item Two ``soft dressing'' unitary matrices  that act solely on the space of soft photons and that depend on  $|a\rangle$ and $|b\rangle$. \footnote{Factorization breaks down for large number of soft quanta, when back-reaction becomes important. However, given that the total emitted energy in soft radiation is much less than $\lambda$ (by a factor of $\alpha$ in electrodynamics and $E^2/\mpl^2$ in gravity), the probability for large back-reaction is negligible and it vanishes in the limit $\lambda\rightarrow 0$.}
\end{enumerate}

{Explicitly:
\beq
\langle \beta|\langle b| \S |a\rangle |\alpha\rangle = \langle b| \hat \S |a\rangle \langle \beta|\Omega(b)
\Omega^\dagger(a)|\alpha\rangle 
\eeq{factorize}
where $\Omega = \Omega_{\rm ph} \Omega_{\rm gr}$; the photon soft factor is given by
\be
\begin{split}
\Omega_{\rm ph}(a) \equiv & \exp\Big(i\int^\lambda \frac{d^3\k}{(2\pi)^3 2|\k|} 
\\
&\sum_s a_{\rm ph}(\k,s)
\ep^*_\mu(s,\k)J^\mu(-|\k|,-\k)
+h.c.\Big),
\end{split}
\ee
$a_{\rm ph}(\k,s)$ is the ladder operator for the free photon field and
\be
J^\mu(|\k|,\k) = -i \sum_{i\in a} \frac{Q_i p_i^\mu}{p_i\cdot k},\qquad\text{with}\quad k^\mu = (|\k|,\k).
\ee
The graviton soft factor is
\be
\begin{split}
\Omega_{\rm gr}(a) \equiv \exp\Big(i\int^\lambda &\frac{d^3\k}{(2\pi)^3 2|\k|}
\sum_s a_{\rm gr}(\k,s)
\ep^*_{\mu\nu}(s,\k) 
\\
&T^{\mu\nu}(-|\k|,-\k)
+h.c.\Big),
\end{split}
\ee
$a_{\rm gr}(\k,s)$ is the ladder operator for the free graviton field and 
\be
T^{\mu\nu}(|\k|,\k) = -i\frac{\kappa}{2}\sum_{i\in a} \frac{p_i^\mu p_i^\nu}{p_i\cdot k}.
\ee
To verify \eqref{factorize} note that Weinberg formula for the emission of multiple soft photons/gravitons is of the form
\be\label{s}
\S_{b,\beta;a,\alpha} = F_{b,\beta;a,\alpha} \S_{b,0;a,0},
\ee
where 
\be
F_{b,\beta;a,\alpha}
= \frac{\langle \beta|\Omega(b)\Omega^\dagger(a)|\alpha\rangle}{ \langle 0|\Omega(b)\Omega^\dagger(a)|0\rangle }.
\ee
So we define
\be\label{shat}
\hat \S_{b;a} \equiv \frac{\S_{b,0;a,0}}{\langle 0|\Omega(b)\Omega^\dagger(a)|0\rangle },
\ee
in terms of which the connected S-matrix reads as \eqref{factorize}. Note that $\hat\S$ is by construction independent of soft states. Note also that dividing by the vacuum expectation value $\langle 0|\Omega(b)\Omega^\dagger(a)|0\rangle$ in \eqref{shat} cancels the IR divergences in $\S_{b,0;a,0}$.

The same techniques developed in \cite{Strominger_graviton,Strominger_photon,Strominger_QED} to establish the 
equivalence of super-translation and large $U(1)$ conservation laws with Weinberg soft formulas, can be  used to show 
that in massless QED  
\be\label{commut}
[\Q_S , \Omega(a)] = \Omega(a)\sum_i  Q_i \vep(\hat \p_i) ,
\ee
and
\be\label{eigen}
\Q_H a_{p_i,\zeta_i}^\dagger|0\rangle = - Q_i \vep(\hat \p_i) a_{p_i,\zeta_i}^\dagger |0\rangle,
\ee
and as a result
\be
\Q^{I^+} \S = \S \Q^{I^-} = \sum_{a,b} |b\rangle\langle a| \langle b |\hat \S|a\rangle \Omega(b) \Q_S \Omega^\dagger(a)
\ee
for all large $U(1)$ charges. Here we used the fact that after antipodal matching of $\vep(z,\bar z)$ on $I^+$ and $I^-$, $\Q_S^{I^+}$ and $\Q_S^{I^-}$ are given by the same expressions in the Fock space of photons. Similar results hold for massive QED as well as gravitational scattering. 

Conversely, the independence of $\hat{\S}$ --defined as $\S$ modulo the soft factors $\Omega$-- from soft photon (or soft graviton) operators also  follows directly from conservation of the current $J=\star d (\vep \star F)$. To prove that, it 
suffices to consider parameters $\vep$ that 
depend on the null coordinates $u,v$ as $\vep_\omega(u,z,\bar{z})=\exp(i\omega u) \eta(z,\bar{z})$ on $I^+$ and
$\vep_\omega(v,z,\bar{z})=\exp(i\omega v) \eta(z,\bar{z})$ on $I^-$. Equation~\eqref{m1} becomes 
\be\label{mm1}
\Q_\omega= \int _{I^+} d (\vep_\omega \star F) = \int_{I^+} \hat{d} \vep_\omega \wedge \star F +  
\int_{I^+} \vep_\omega d \star F + \int_{I^+} du \partial_u\vep_\omega \wedge \star F .
\ee
On $I^-$ a similar equation holds. 

The last term in eq.~\eqref{mm1} vanishes in the limits $\omega\rightarrow 0^\pm$. This can be proven using
$|\int_{I^+} du \partial_u\vep_\omega \wedge \star F|= |\oint_{S^2} \omega \eta \tilde{F}_{ur}|$. The Fourer transform 
$\tilde{F}_{ur} $ of the field strength $F_{ur}$  is $L^2$ since 
$\oint_{S^2}\int d\omega |\tilde{F}_{ur}|^2 =\oint_{S^2}\int dt |{F}_{ur}|^2$, by Parseval's identity, and  
$\oint_{S^2}\int dt |{F}_{ur}|^2 \leq \oint_{S^2}\int dt {\cal H}<\infty$. Here $\cal H$ is the EM energy density.

Conservation of $\Q_\omega$ thus implies, after using \eqref{commut} and \eqref{eigen} and with obvious notation
\be\label{mm2}
\lim_{\omega\rightarrow 0^\pm} [\Q_{\omega\, S} ,\hat{\S}]=0.
\ee
Now it suffices to recall~\cite{Strominger_photon,Strominger_QED} that $\lim_{\omega\rightarrow 0^+}\Q_{\omega\, S}$
creates a soft photon, while $\lim_{\omega\rightarrow 0^-}\Q_{\omega\, S}$ annihilates it, to conclude that
$\hat{\S}$ commutes with {\em all} soft photon creation and annihilation operators. By Shur's lemma this means that
$\hat{\S}$ is a constant on Fock space of the soft photons, since that space is an irreducible representation of the 
canonical commutation relations.\footnote{Refs.~\cite{Strominger_photon,Strominger_QED} impose the weaker requirement that $\lim_{\omega\to 0^+} \frac{1}{2}(\Q_{\omega}+\Q_{-\omega})$ commute with the S-matrix. However, to derive the soft theorem one has to use an additional identity, {valid on a dense subset of states}:
\be
\lim_{\omega\to0} a_{\rm ph}(\omega \hat x,+) \S = - \S a_{\rm ph}^\dagger (\omega \hat x,-).
\ee
This identity follows from the absence of monopole interaction \cite{Mirbabayi}. Combined together they imply that both $\lim_{\omega\to 0^\pm} \Q_{\omega}$ commute with $\S$.}

We introduce now a new basis of scattering states, obtained from the old ones by dressing the hard 
particles as~\cite{chung}
\be\label{dressed}
||a,\alpha \rangle\rangle = \Omega(a) |a\rangle |\alpha\rangle, 
\quad |a\rangle \in \H^-_h,\quad |\alpha\rangle \in \H^-_s.
\ee
In this basis the soft part $\alpha$ evolves trivially and all dynamics is in the hard part:
\be\label{SS}
\langle\langle b,\beta|| \S || a,\alpha\rangle\rangle = \langle b|\hat\S |a\rangle \langle \beta|\alpha \rangle.
\ee
Working in this basis makes it clear that during the Hawking evaporation (1) super-translation and large $U(1)$ symmetries put no constraint on the hard radiation, and (2) for a big black hole early and late hard quanta are separately accompanied by their 
own soft radiation $\Omega(a_{\rm early})$ and $\Omega(a_{\rm late})$. No information is carried over from the early stage
 of evaporation to the later period. In other words, the soft dynamics decomposes into superselection sectors that
never mix during time evolution.

\section{Additional Remarks}

The factorized S-matrix (\ref{SS}) also explains why {neither} electromagnetic {nor} gravitational memory can be regarded as black hole hair. Imagine two black holes of equal mass $M$; one of them formed by colliding two high energy photons along the $x$ axis $|p_x,-p_x\rangle$ and the other by the same collision along the $y$ axis $|p_y,-p_y\rangle$. According to \cite{Strominger_memory} this directional information can be retrieved by looking at the soft gravitational emission $|\alpha\rangle$ from the formation process. Thus it seems that less information is needed to be stored in black hole for the whole process of black hole formation and evaporation to be unitary.

However, this argument ignores the possibility of having soft incoming radiation. Once that is included, for any observed 
gravitational memory $|\alpha\rangle$ the kinematics of hard incoming states remains completely undetermined. In 
particular, the two initial states $||p_x,-p_x,\alpha\rangle\rangle$ and $||p_y,-p_y,\alpha\rangle\rangle$ produce mass-$M$ 
black holes with identical gravitational memories $|\alpha\rangle$.

A generic state is an entangled superposition of soft and hard states 
\be\label{m-ent}
|V\rangle=\sum_{a\,\alpha}C(a,\alpha) ||a,\alpha\rangle\rangle, \qquad C(a,\alpha)\in \mathbb{C},
\ee
but any such entanglement cannot be used to extract information
on the state $|V\rangle$ using operators that act only on hard modes. Specifically, a large $U(1)$ transformation is a 
unitary operator $U$ that, in the new basis of dressed states $\{ ||a,\alpha\rangle\rangle \}$, acts only on soft states; so, it does not affect the matrix elements of any operator $O$ that 
depends only on hard quanta because
$U^{\dagger} O U=O$. In particular, $\langle V | O |V \rangle= \langle V' | O | V'\rangle$, $|V'\rangle=U|V\rangle$. The 
S-matrix, seems at first sight to mix hard and soft modes, but in the  basis of dressed states
$|| a,\alpha \rangle\rangle$, we have shown 
that it also factorizes into the product of an operator acting only on hard modes plus
the identity operator acting on soft modes in the basis $\{ || a,\alpha \rangle\rangle \} $.

{It is worth expanding on the last remark and come to the original analogy with Lorentz boosts, to} ask what 
is the fundamental difference between conservation laws associated to super-translations (and large $U(1)$'s) and 
momentum conservation. Note that after the emission of early quanta, the remaining black hole is not just boosted to cancel 
the net momentum transferred to the early radiation. Due to the soft graviton/photon radiation, it is also immersed in a 
vacuum with a different metric, {and a different}  $A_\mu$ configuration. Inside the light cone created by the early soft 
radiation this is a pure gauge configuration which can be generated from the vacuum by a large gauge transformation. Let 
us focus for simplicity on the electromagnetic case and study the action of the generator of the large $U(1)$ 
{transformations} as in \cite{Hawking}
\be\label{QM}
|\tilde M\rangle = \Q |M\rangle.
\ee
The conservation of $\Q$ implies that $|\tilde M\rangle$ evaporates into $\Q|X\rangle$. However, we should now include the soft radiation in $|X\rangle$:
\be
\Q |X\rangle = \sum_b \hat \S_{M\to b} \Q ||b,0\rangle\rangle = \sum_b \hat \S_{M\to b} ||b,\alpha\rangle\rangle
\ee
where we used \eqref{commut}, \eqref{eigen}, and their analogs, and we defined:
\be 
|\alpha\rangle = \Q_S|0\rangle.
\ee
This is an exactly zero-frequency photon. In reality $\Q_S$ is IR regulated by the distance that the
early radiation has traveled until the detection of late quanta.\footnote{Continuation of asymptotic charges to finite distance was discussed in \cite{Mirbabayi}.} This distance is much larger than the box over which the late-time detector makes measurements. Hence, the late-time observer has no way of distinguishing $|\alpha\rangle$ from $|0\rangle$. Therefore, unlike a boost, which transforms late Hawking quanta but leaves the vacuum invariant, spontaneously broken super-translations and large $U(1)$'s leave measurable Hawking quanta invariant and {merely} unobservably transform the vacuum.

It is amusing to notice that here the factorization of soft photons into superselection sectors {is crucial to explain why the information paradox persists}, while in the context of the ``baby universe'' picture of black hole evaporation, advocated in~\cite{polchinski-strominger,strominger-lh}, a superselection-sector factorization was crucial to that  proposal for solving the puzzle. 

{Various constructions of different types of soft hair have been proposed in the literature.
In particular, the hair defined in~\cite{cd1,*cd2,*cd3} are effects due to the finite number of quanta
involved in a ``corpuscular'' description of black holes. The hair studied in~\cite{cd5,*cd6,*cd7},
as well as those in~\cite{s-j}, are associated to symmetries of the horizon. The soft hair 
considered in this paper are instead specifically only those connected with the large asymptotic 
$U(1)$ (or BMS) symmetries considered in~\cite{Hawking}.}

{A deep, extensive analysis of QED, which uses dressed states similar to~(\ref{dressed}) and includes a construction of the
S-matrix, was given in a remarkable series of paper
by Kibble~\cite{kibble1,*kibble2,*kibble3,*kibble4}. We thank A. Schwimmer for making us aware of those paper. 
A similar construction of the S-matrix using coherent states was  proposed also in~\cite{fk}. 
After this paper was posted to the archives, we received the draft of a manuscript~\cite{Gabai} that 
independently arrives to conclusions similar to ours. We thank the authors for sending it to us
prior to publication.}

\section*{Acknowledgements}

It is a pleasure to thank D. Harlow, M. Kleban, J. Maldacena, A. Maloney, A. Schwimmer, A. Strominger, S. Yankielowicz
and A. Zhiboedov for useful discussions and comments on the paper. MM is supported in part by NSF grants PHY-1314311 and PHY-0855425. MP would like to thank Kavli IPMU, Tokyo, for its kind hospitality during completion of this paper. MP is supported in part by NSF grant PHY-1316452.  

\bibliography{bibhair}

\end{document}